\documentclass[onecolumn]{article}
\usepackage{ascmac} 
\usepackage{graphicx}
\usepackage{latexsym}
\usepackage{cite}
\usepackage{amsmath}
\usepackage[psamsfonts]{amssymb}
\usepackage{amsfonts}
\usepackage{arydshln}
\usepackage{longtable}
\usepackage{threeparttable}
\usepackage{dashbox}
\usepackage{multirow}
\usepackage{accents}
\usepackage{fancybox,ascmac}

\def\vec#1{\mbox{\boldmath$#1$}}

\def\vv#1{\mbox{\boldmath$#1$}}
\def\bb{\begin{equation}\begin{aligned}}
\def\ee{\end{aligned}\end{equation}}

\def\ftwo{$\mathbb{F}_2$}

\def\z0{$\mathbb{Z}$}

\def\ii{{I\nobreak\hspace{-0.05cm}I}}
\def\iii{{I\nobreak\hspace{-0.05cm}I\nobreak\hspace{-0.05cm}I}}

\def\xi{{X\nobreak\hspace{-0.05cm}I}}

\title{
Constructions of A Large Class of Optimum Constant Weight Codes over \ftwo
}
\author{
  Masao KASAHARA           
  \thanks{
   Research Institute for Science and Engineering, Waseda University.
   Research and Development Initiative, Chuo University.
    kasahara@ogu.ac.jp
  }
  \and
  Shigeichi HIRASAWA       
  \thanks{
   Research Institute for Science and Engineering, Waseda University.
    hira@waseda.jp
  }
}

\date{}

\begin{document}
\maketitle

\section*{Abstract}
A new method of constructing optimum constant weight codes over \ftwo ~based on a generalized 
$(\vv u, \vv u + \vv v)$ construction\cite{bib:3}$\sim$\cite{bib:4} is presented.
We present a new method of constructing superimposed code $C_{(s_1,s_2,\cdots,s_I)}^{(h_1, h_2, \cdots, h_I)}$ bound.
and presented a large class of optimum constant weight codes over \ftwo ~that meet the bound due to Brouwer and Verhoeff\cite{bib:1},
which will be referred to as BV .
We present large classes of optimum constant weight codes over \ftwo ~for $k=2$ and $k=3$ for $n \leqq 128$.
We also present optimum constant weight codes over \ftwo ~that meet the BV bound \cite{bib:1} for $k=2,3,4,5$ and 6, for $n \leqq 128$. 
The authors would like to present the following conjectures : \\
\begin{tabular}{lcl}
$C_I$       &:& $C_{(s_1)}^{(h_1)}$ presented in this paper yields the optimum constant weight codes for the code-length\\
            & & $n=3h_1$, number of information symbols $k=2$ and minimum distance $d=2h_1$ for any positive\\
            & & integer $h_1$. \\
$C_\ii$     &:& $C_{(s_1)}^{(h_1)}$ yields the optimum constant weight codes at $n=7h_1, k=3$ and $d=4h_1$ for any $h_1$. \\
$C_\iii$    &:& Code $C_{(s_1,s_2,\cdots,s_I)}^{(h_1, h_2, \cdots, h_I)}$ yields the optimum constant weight codes of length $n=2^{k+1}-2$, and \\
            & &minimum distance $d=2^{k}$ for any number of information symbols $k\geq 3$.\\
\end{tabular}
\section*{keyword}
Optimum code, Constant weight code, Brouwer$\cdot$Verhoef bound, $(\vv u, \vv u + \vv v)$ construction

\maketitle

\section{Introduction} 

Extensive studies have been made of the construction of efficient codes over \ftwo\cite{bib:3}$\sim$\cite{bib:4}. 
In addition to cooperation with these studies the various bounds on the minimum distance of linear 
and non-linear codes have also been reported\cite{bib:1},\cite{bib:5}. 
In the 1970's the authors proposed the various classes of efficient codes on both random and compound channels 
based on the product code, the BCH code and the $(\vv u, \vv u + \vv v)$ construction\cite{bib:3}$\sim$\cite{bib:4}. 
A large number of constructed codes over \ftwo ~are listed in Table of the best known codes 
compiled by MacWilliams and Sloane\cite{bib:3}. 
Hereafter we shall denote the code by $(n, k, d)$ code, 
where $n, k$ and $d$ denote code-length, number of information symbols and minimum distance respectively.

In this paper we propose a new method of constructing a large class of optimum constant weight codes over \ftwo\cite{bib:3} 
based on a generalized $(\vv u, \vv u + \vv v)$ construction.

We have presented a new method of constructing superimposed code $C_{(s_1,s_2,\cdots,s_I)}^{(h_1, h_2, \cdots, h_I)}$ 
and presented a large class of optimum constant weight codes.

We present large classes of optimum constant weight codes that meet the bound due to Brouwer and Verhoeff\cite{bib:1} (BV bound) 
for $k=2$, $k=3$ for $n\leqq 128$.

We also present optimum constant weight codes that meet BV bound for $k=2,3,4,5$ and 6, for $n \leqq 128$.

\section{Codes $C_2$ and $C_3$} 
\subsection{Code $C_2$} 

In Fig.1 we show the principle of constructing superimposed code, over \ftwo.
This construction, $(\vv u, \vv u + \vv v)$ construction, 
yields the Reed-Muller codes and various other classes of efficient codes. 
In conventional constructions, the code $\{ \vv v \}$ is chosen so that it may be capable of correcting twice as many errors, 
compared with the code $\{ \vv u \}$.
In the followings, these codes $\{ \vv u \}$ and $\{ \vv v \}$ will be referred to as $\vv u$ 
and code $\vv v$ code respectively. 
In this paper the superimposed code shown in Fig.1 will be referred to as code $C_2$.
Let the code word of the code $\{ C \}$ be denoted by the polynomial $C(x)$ and by the vector $\vv C$, over \ftwo, as follows : 
\begin{eqnarray}
  C(x)  = u(x) + u(x) x^{\frac{n}{2}} + v(x)x^{\frac{n}{2}}
\end{eqnarray}
and
\begin{eqnarray}
  \vv C = (\vv u, \vv u + \vv v).
\end{eqnarray}

\setcounter{figure}{0}
\begin{figure}[h]
\begin{center}
\includegraphics[scale=0.5]{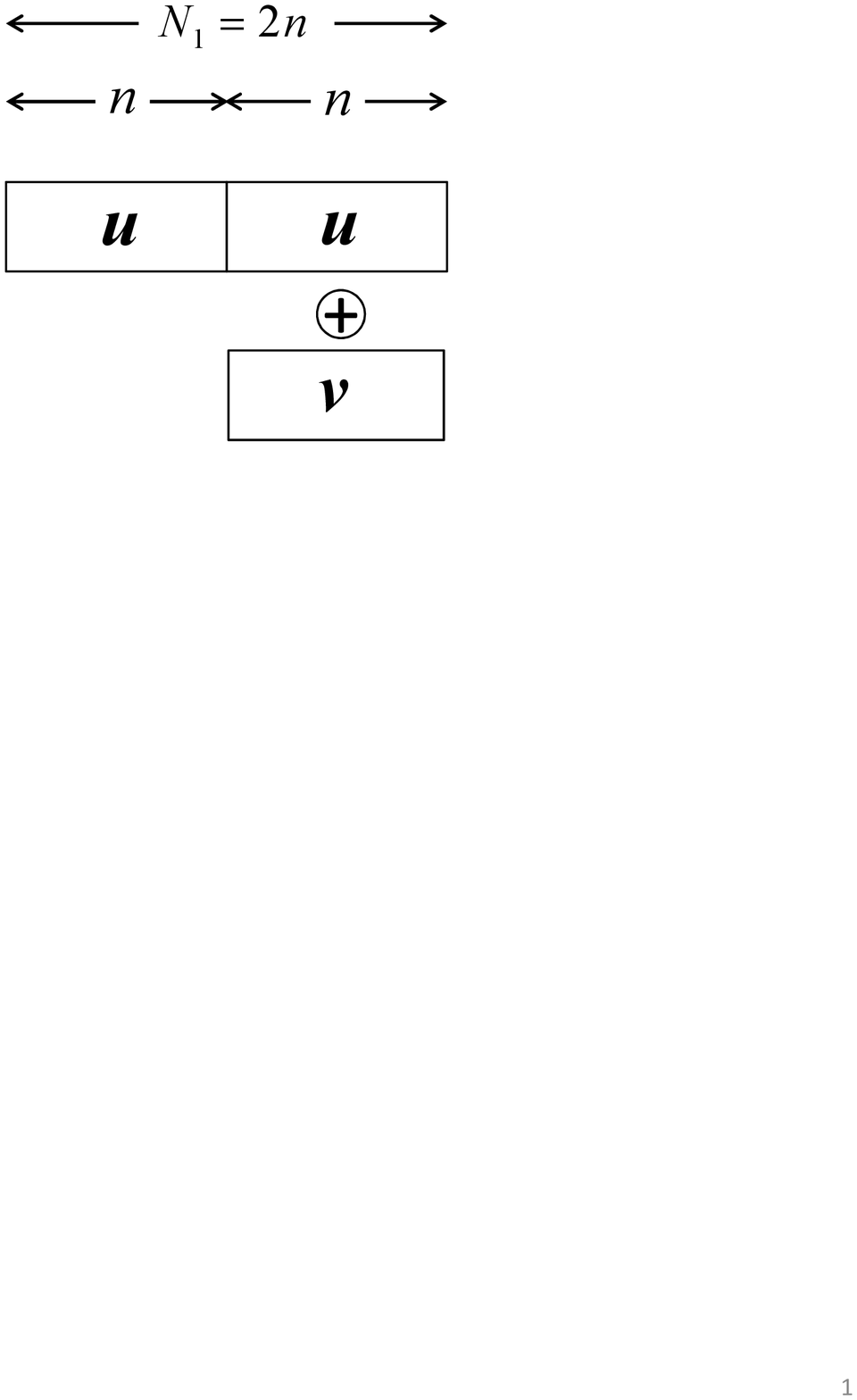}
\end{center}
\label{fig:fig1}
\caption{Structure of code, $C_2$.}
\end{figure}

Similarly let the channel errors be denoted by the polynomial $e(x)$ and by the vector $\vv e$ over \ftwo, as follows: 
\begin{eqnarray}
  e(x)  = e_{\rm L}(x) + e_{\rm R}(x) x^{\frac{n}{2}},
\end{eqnarray}
\begin{eqnarray}
  \vv e = (\vv e_{\rm L}, \vv e_{\rm R}),
\end{eqnarray}
where $e_{\rm L}(x)$ denotes the random errors that occur on the code word $u(x)$ 
and $e_{\rm R}(x)$, on the code word of $u(x) + v(x)$. 
The received word over $\mathbb{F}_2$ can then be represented by
\begin{eqnarray}
  \vv r = (\vv r_{\rm L}, \vv r_{\rm R}),
\end{eqnarray}
where
\begin{eqnarray}
  \vv r_{\rm L} = \vv u + \vv e_{\rm L}
\end{eqnarray}
and
\begin{eqnarray}
  \vv r_{\rm R} = \vv u + \vv v + \vv e_{\rm R}.
\end{eqnarray}

From the received word $\vv r$, we first generate an erroneous version of the code word $\vv v$ as follows :
\begin{eqnarray}
  \vv r_{\rm L} + \vv r_{\rm R} = \vv v + \vv e_{\rm R} + \vv e_{\rm L}.
\end{eqnarray}

Let the Hamming weight of the vector $\vv v$ be denoted by $w(\vv v)$ and the minimum distance of the code $\{ \vv v \}$, 
$d_{\vv v}$, then the code word $\vv v$ can be decoded correctly, with the conventional bounded distance decoding algorithm, 
if and only if the following relation holds :
\begin{eqnarray}
  w(\vv e_{{\rm L}} + \vv e_{{\rm R}}) \leq \left\lfloor \frac{d_{\vv v} - 1}{2} \right\rfloor,
\end{eqnarray}
where $\lfloor x \rfloor$ denotes the largest integer less than or equal to $x$.
Subtracting the decoded $\vv v$ from the received word $\vv r$, we obtain the following $\vv r_{D}$ : 
\begin{eqnarray}
  \vv r_D = (\vv u + \vv e_{{\rm L}}, \vv u + \vv e_{{\rm R}}).
\end{eqnarray}

Again with the minimum distance decoding algorithms, 
$\vv r_D$ can be decoded successfully if and only if the minimum distance, $d_{\vv u}$, of the code
$\{ (\vv u, \vv u) \}$ satisfies the following relation:
\begin{eqnarray}
  w\{ (\vv e_{\rm R}, \vv e_{\rm L}) \} \leq d_{\vv u} - 1.
\end{eqnarray}

\subsection{Code $C_3$} 
By further cascading code word $\vv u$ in $s$ steps, we can construct code $C_s$.
Assuming that $s = 3$, we shall present here several efficient codes 
that satisfy the minimum distance bound of the codes listed in Ref.\cite{bib:1}, BV bound.
In Fig.2, we let\\
\begin{tabular}{rcl}
  $\{ \vv u \}$ & : & Maximum-period sequence of length $n$, M-sequence\cite{bib:3}.\\
  $\{ \vv v \}$ & : & $\{0^{(2n)}, 1^{(2n)}\};n = 2^{\nu} - 1$, \\
\end{tabular}\\
where $a^{(n)}$ denotes all $a$'s vector of length $n$.

Let us present several examples of optimum codes among code $C_3$, in Table 1.
In the table $d_{BV}$ denotes the minimum distance bound due to Brouwer and Verhoeff, referred to as BV bound.

However it should be noted that the codes given in Table 1 are not the constant weight codes, although M-sequence 
is a member of constant weight codes.

\vspace{2em}

\begin{figure}[h]
\begin{center}
\includegraphics[scale=0.5]{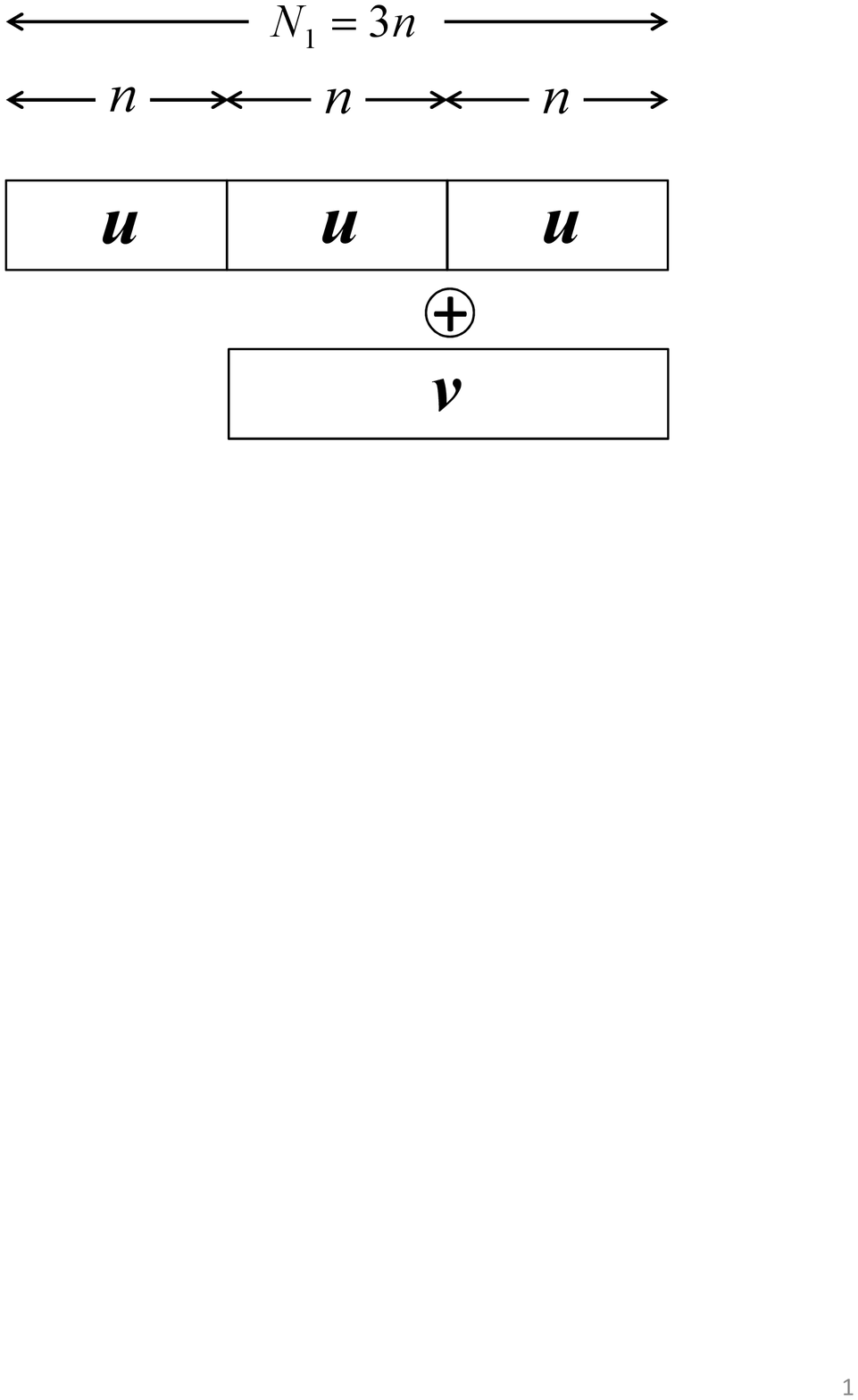}
\end{center}
\label{fig:fig2}
\caption{Structure of $C_3$.}
\end{figure}
\begin{table}[h]
\begin{center}

\caption{Examples of code $C_3$.}
\scalebox{1.20}{
\begin{tabular}{|c|c|c|c|c|c|} \hline
  $\nu$ & $n$ & $N_1$ & $k$ & $d$    & $d_{BV}$  \\ \hline
  2     & 3   & 9     & 3   & 4      & $4$  \\ \hline
  3     & 7   & 21    & 4   & 10     & $10$ \\ \hline
  4     & 15  & 45    & 5   & 22     & $22$ \\ \hline
  5     & 31  & 93    & 6   & 46     & $46$ \\ \hline
\end{tabular}
}
\end{center}
\end{table}

\newpage
\section{Codes $C_{(s_1)}^{(h_1)}$} 

Let us present the code $C^{(h_1)}_{(s_1)}$ by modifying the code $C_2$. 
We show the modified code $C^{(h_1)}_{(3)}$ given in Fig.3 as an example. 

\vspace*{1em}

\noindent {\bf Example 1: $C_{(s_1)}^{(h_1)}$ for $s_1=2$, $h_1=1$, $k=2; C_{(2)}^{(1)}(k=2)$.}

The simplest version of $C_{(s_1)}^{(h_1)}$ for $k=2$, $C_{(2)}^{(1)}(k=2)$, can be constructed based on the following codes:
\bb
 \{ \vv u \} = \{ 0,1 \}
\ee
and 
\bb
 \{ \vv v \} = \{ 00,11 \}.
\ee

The code $C_{(2)}^{(1)} (k=2)$ is shown below :

\begin{table}[!h]
\begin{center}
\scalebox{1.20}{
\begin{tabular}{|cc|c|} 
 \multicolumn{3}{c}{Code $C_{(2)}^{(1)}(k=2)$.} \\ \hline
 \multicolumn{2}{|c|}{message}   & \raisebox{-1.5ex}[0cm][0cm]{code word}  \\ 
  $m_1$ & $m_2$   &            \\ \hline
   0    &  0      &  0 0 0  \\
   0    &  1      &  0 1 1  \\
   1    &  0      &  1 1 0  \\
   1    &  1      &  1 0 1  \\ \hline
\end{tabular}
}
\label{tab:ex:C3}
\end{center}
\end{table}

\begin{figure}[h]
\begin{center}
\includegraphics[scale=0.5]{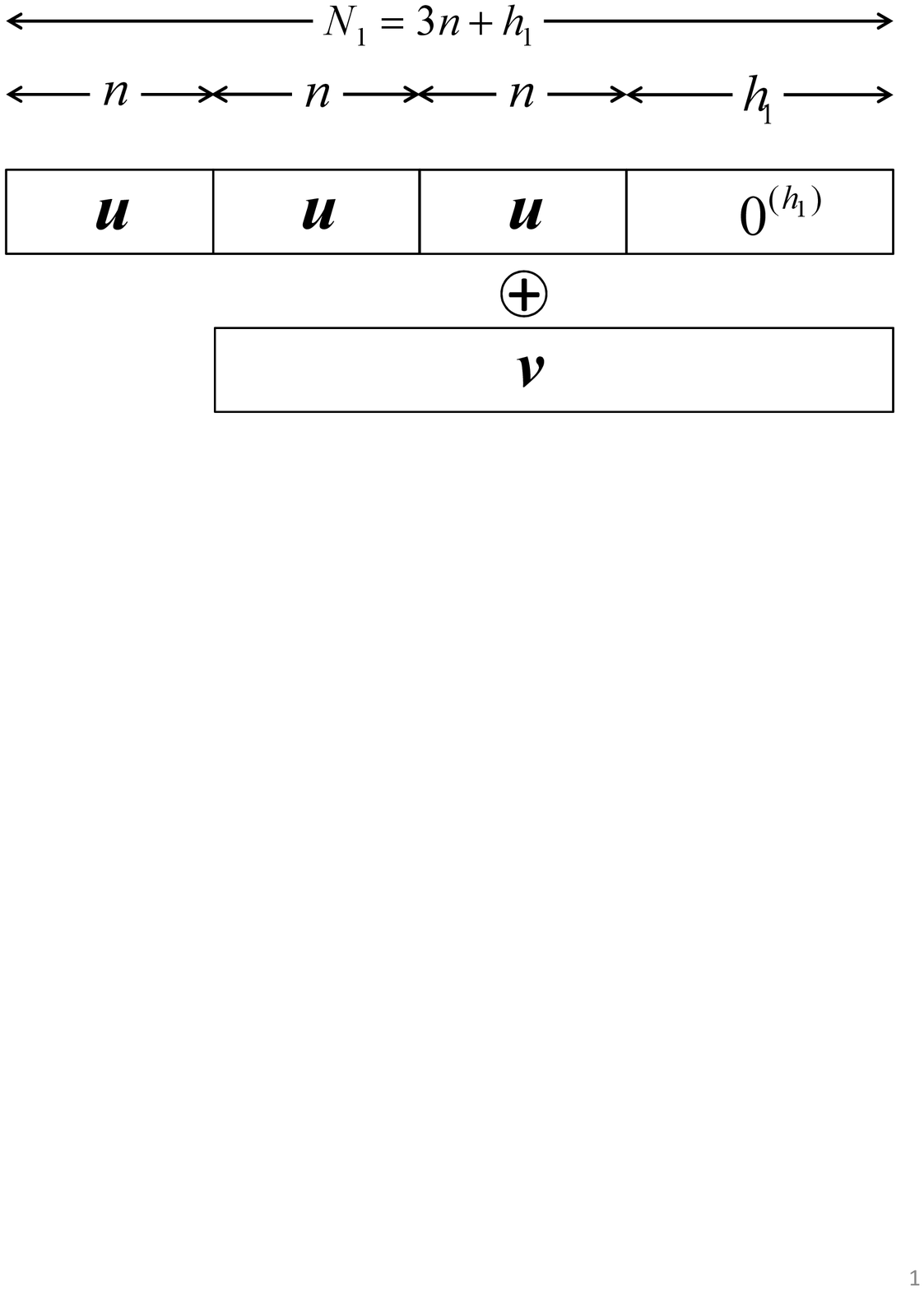}
\end{center}
\label{fig:fig4}
\caption{Code $C_{(3)}^{(h_1)}$}
\end{figure}

We see that $C_{(2)}^{(1)} (k=2)$ is a constant weight code that meets BV bound.

\vspace*{1em}

In the followings, let us present the modified code $C_{(s_1)}^{(h_1)}$ for $h_1=2,s_1=4$ and $k=2$, $C_{(4)}^{(2)}(k=2)$.

Let the code $\{ \vv u \}$ and $\{ \vv v \}$ be
\begin{eqnarray}
  \{ \vv u \} = \{ 0, 1\}
\end{eqnarray}
and 
\begin{eqnarray}
  \{ \vv v \} = \{ 0^{(J)}, 1^{(J)} \},
\end{eqnarray}
where $J$ is chosen as $s_1$.

\vspace*{1em}

In the followings let us present an example of optimum constant weight code for $k=2$. 

\vspace*{1em}

\noindent {\bf Example 2: $C_{(s_1)}^{(h_1)}$ for $s_1=4, J=s_1$, $h_1=2; C_{(4)}^{(2)}$} 

As $J$ is $J=s_1=4$, the code words $\{ \vv u \}$ and $\{ \vv v \}$ are $\{ \vv u \}=\{ 0,1 \}$ 
and $\{ \vv v \}=\{ 0000,1111 \}$, yielding the code words, as shown below.
We see that the code is (6,2.4) constant weight code.

\begin{table}[h]
\begin{center}
\scalebox{1.2}{
\begin{tabular}{|cc|c|} \hline
 \multicolumn{2}{|c|}{message}   & \raisebox{-1.5ex}[0cm][0cm]{code word}  \\ 
  $m_1$ & $m_2$  &                 \\ \hline
   0    &   0    &  0 0 0 0 0 0    \\
   0    &   1    &  0 0 1 1 1 1    \\
   1    &   0    &  1 1 1 1 0 0    \\
   1    &   1    &  1 1 0 0 1 1    \\ \hline
\end{tabular}
}
\label{tab:ex:C3}
\end{center}
\end{table}

\vspace*{1em}

\noindent {\bf Theorem 1:} 
The code $C_{(s_1)}^{(h_1)} (k=2)$ is a constant weight code when $h_1$ satisfies $h_1=\frac{s_1}{2}$.

\vspace*{1em}

\noindent {\bf Proof:} 
Let the code word be denoted by $\vv F_i$.
For any pair $(\vv F_i, \vv F_j) (i \neq j)$, the weight of $\vv F_i+\vv F_j, w_{ij}$, is given as shown below for $h_1=\frac{s_1}{2}$ :

\begin{table}[h]
\begin{center}
\scalebox{1.2}{
\begin{tabular}{ccl} 
  $m_1$ & $m_2$  &  $w_{ij}$ \\
   0    &   0    &  0       \\
   0    &   1    &  $s_1$     \\
   1    &   0    &  $s_1$     \\
   1    &   1    &  $\frac{s_1}{2}+h_1=s_1$  \\ 
\end{tabular}
}
\end{center}
\end{table}
\noindent , yielding the proof. \hfill $\Box$

In Table 2, we present the examples of optimum constant weight codes for $k=2$.
We see that, for $k=2$, the constant weight codes exist for any $s_1=$even number for $N_1 \leqq 126$.
It is strongly conjectured that code $C_{(s_1)}^{(s_1/2)} (k=2)$ is a constant weight code that meets BV bound, 
for any code length $N_1 = 3h_1$.

\begin{table}[h]
\begin{center}
\addtocounter{table}{0}
\caption{Constant weight code, $C_{(s_1)}^{(h_1)}(k=2)$.}
\scalebox{1.2}{
\begin{tabular}{|c|c|c|c|c|} \hline
  $s_1$  & $h_1$ & $N_1$   & $d$  & $d_{BV}$  \\ \hline
   2     &  1    &    3    &  2   &  2        \\ \hline
   4     &  2    &    6    &  4   &  4        \\ \hline
   6     &  3    &    9    &  6   &  6        \\ \hline
   8     &  4    &   12    &  8   &  8        \\ \hline
         &       & \vdots  &      &           \\ \hline
  78     & 39    &  117    & 78   & 78        \\ \hline
  80     & 40    &  120    & 80   & 80        \\ \hline
  82     & 41    &  123    & 82   & 82        \\ \hline
  84     & 42    &  126    & 84   & 84        \\ \hline
\end{tabular}
}
\label{tab:ex:C3}
\end{center}
\end{table}

\vspace{10em}

\section{Code $C^{(h_1, h_2, \cdots, h_I)}_{(s_1, s_2, \cdots, s_I)}$} 

By superimposing the superimposed codes iteratively, a new class of codes are constructed.
We shall refer to the constructed code as $C^{(h_1, h_2, \cdots , h_I)}_{(s_1, s_2, \cdots, s_I)}$.

\vspace{2em}

\begin{figure}[h]
\begin{center}
\includegraphics[scale=0.5]{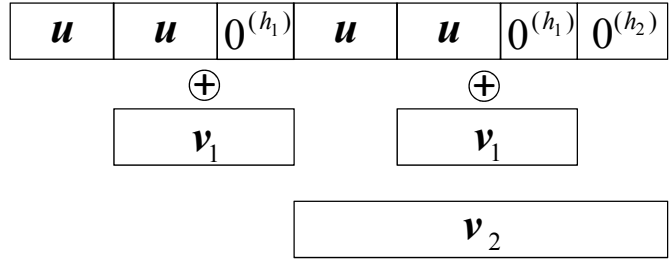}
\end{center}
\label{fig:fig5}
\caption{\small Structure of $C_{(2,2)}^{(h_1,h_2)}$.}
\end{figure}

In Fig.\ref{fig:fig5}, we show an example of code $C^{(h_1, h_2)}_{(s_1,s_2)}$ based on $C^{(h_1)}_{(s_1)}$,
where $(\vv u, \vv u * 0^{(h_1)}+ \vv v_1)$ is the code word of $C^{(h_1)}_{(s_1)}$.

The code word of $C^{(h_1, h_2)}_{(s_1,s_2)}$, $\vv w^{(h_1, h_2)}$, can be represented by 
\begin{eqnarray}
  \vv w^{(h_1, h_2)} =  &(\vv u, \vv u * 0^{(h_1)} + \vv v_1, \\ \nonumber
                            &(\vv u, \vv u * 0^{(h_1)} + \vv v_1 ) + \vv v_2),
\end{eqnarray}
where $*$ implies the concatenation.
It is easy to see that when $h_1 = h_2 = 0$, then $C^{(0, 0)}_{(2,2)}$ can be constructed 
by iterating the conventional $(\vv u, \vv u + \vv v)$ construction.

Thus the method presented in this paper could be referred to as a generalized $(\vv u, \vv u + \vv v)$ construction 
or $g \cdot (\vv u, \vv u + \vv v)$ construction.

In the followings let us present several examples of optimum constant weight codes, 
among $C_{(s_1,s_2,\cdots,s_I)}^{(h_1,h_2,\cdots,h_I)}$.

\vspace*{1em}

\noindent {\bf Example 3: $C_{(s_1,s_2)}^{(h_1,h_2)}$ for $s_1=2, s_2=2, h_1=1$ and $h_2=1, C_{(2,2)}^{(1,1)} (k=3)$.}

Let $J$ be 4, the code words $\{ \vv u \}$ and $\{ \vv v \}$ are $\{ (000), (011), (110), (101)\}$ and $\{(0000), (1111)\}$.

The constructed $C_{(2,1)}^{(1,1)} (k=3)$ is shown below.

\begin{table}[h]
\begin{center}
\scalebox{1.2}{
\begin{tabular}{|ccc|c|} \hline
 \multicolumn{3}{|c|}{message}   & \raisebox{-1.5ex}[0cm][0cm]{code word}  \\ 
  $m_1$ & $m_2$ & $m_3$ &                 \\ \hline
   0    &   0   &   0   &  0 0 0 0 0 0 0    \\
   0    &   0   &   1   &  0 0 0 1 1 1 1    \\
   0    &   1   &   0   &  0 1 1 0 1 1 0    \\
   0    &   1   &   1   &  0 1 1 1 0 0 1    \\
   1    &   0   &   0   &  1 1 0 1 1 0 0    \\
   1    &   0   &   1   &  1 1 0 0 0 1 1    \\
   1    &   1   &   0   &  1 0 1 1 0 1 0    \\
   1    &   1   &   1   &  1 0 1 0 1 0 1    \\ \hline
\end{tabular}
}
\label{tab:ex:C3}
\end{center}
\end{table}

Let $J$ be $J=\frac{3s_2}{2}+h_1=8$, the code words $\{ \vv u \}$ and $\{ \vv v \}$ are $\{(000), (011), (110), (101)\}$ and 
$\{(00000000), (11111111)\}$.

As shown below, $C_{(s_1,s_2)}^{(h_1,h_2)}$ for $h_1=1, h_2=2, s_1=2$ and $s_2=4$ constructed  based on $C_{(2)}^{(1)} (k=2)$
is a $(14,3,8)$ constant weight code.
It should be noted that $(14,4,7)$ code obtained by shortening $(15,4,8)$ maximum length code
by 1 bit is not a member of the class of constant weight codes.

\vspace*{1em}

\begin{table}[h]
\begin{center}
\scalebox{1.2}{
\begin{tabular}{|cc|c|c|c|c|} \hline
             $m_1$              &               $m_2$            & $m_3$ &             $\vv u$              & $\vv v$  & code word          \\ \hline
 \raisebox{-1.5ex}[0cm][0cm]{0} & \raisebox{-1.5ex}[0cm][0cm]{0} &   0   & \raisebox{-1.5ex}[0cm][0cm]{000} & 00000000 & 000 000 000 000 00 \\
                                &                                &   1   &                                  & 11111111 & 000 000 111 111 11 \\ \hline
 \raisebox{-1.5ex}[0cm][0cm]{0} & \raisebox{-1.5ex}[0cm][0cm]{1} &   0   & \raisebox{-1.5ex}[0cm][0cm]{011} & 00000000 & 011 011 011 011 00 \\
                                &                                &   1   &                                  & 11111111 & 011 011 100 100 11 \\ \hline
 \raisebox{-1.5ex}[0cm][0cm]{1} & \raisebox{-1.5ex}[0cm][0cm]{0} &   0   & \raisebox{-1.5ex}[0cm][0cm]{110} & 00000000 & 110 110 110 110 00 \\
                                &                                &   1   &                                  & 11111111 & 110 110 001 001 11 \\ \hline
 \raisebox{-1.5ex}[0cm][0cm]{1} & \raisebox{-1.5ex}[0cm][0cm]{1} &   0   & \raisebox{-1.5ex}[0cm][0cm]{101} & 00000000 & 101 101 101 101 00 \\
                                &                                &   1   &                                  & 11111111 & 101 101 010 010 11 \\ \hline
\end{tabular}
}
\label{tab:ex:C3}
\end{center}
\end{table}

\vspace*{1em}

\noindent {\bf Theorem 2:} Code $C_{(s_1,s_2)}^{(h_1,h_2)} (k=3)$ constructed based on $C_{(2)}^{(h_1)} (k=2)$ 
is a constant weight code when $h_1=\frac{s_2}{2}$ and $J= 3 \times \frac{s_2}{2} +h_1 =2s_2$.

\vspace*{1em}

\noindent {\bf Proof:} Referring to Examples 2 and 3, we see that the following relation holds, 
for $h_1=\frac{s_2}{2}$ and $J=s_2$:
\bb
 N_1 = \frac{3s_2}{2}+\frac{3s_2}{2}+h_1 = \frac{7s_2}{2}
\ee
The weight of any non-zero code word is given by $2s_2$ for $\vv v = \vv 0$ or $2 \times \frac{s_2}{2} + \frac{s_2}{2} +h_1 =2s_2$ 
for $\vv v \neq \vv 0$. \hfill $\Box$

From Theorem 2 we see that a large class of optimum constant weight code of $k=3$ exists for $s_1=1$ and any even $s_2$.

Let $\{ \vv u \}$ and $\{ \vv v \}$ be
\begin{eqnarray}
 \{ \vv u \} = C_{(2)}^{(1)} (k=2) = \{ (000), (011), (110), (101)\}
\end{eqnarray}
and
\begin{eqnarray}
  \vv v  = \left\{ 0^{(3s_1/2+h_1)}, 1^{(3s_1/2+h_1)} \right\}.
\end{eqnarray}

\vspace*{1em}

In Table 3 and Table 4 we present several examples of optimum linear constant weight code of three information symbols, 
$C_{(s_1,s_2,\cdots,s_5)}^{(h_1,h_2,\cdots,h_5)}$ for $N_1 \leqq 128$.

\begin{table}[h]
\begin{center}
\addtocounter{table}{0}
\caption{Constant weight code $C_{(2,s_2)}^{(2,h_1)}$ for $k=3$.}
\scalebox{1.2}{
\begin{tabular}{|c|c|c|c|c|c|c|c|c|c|c|} \hline
$s_2$ & $h_1$ & $N_1$ & $d$ & $d_{BV}$ & &$s_2$& $h_1$ & $N_1$ & $d$ & $d_{BV}$  \\ \hline
  2   & 1     & 7   & 4     & 4        & & 20  & 10    & 70  & 40    & 40        \\ \hline
  4   & 2     & 14  & 8     & 8        & & 22  & 11    & 77  & 44    & 44        \\ \hline
  6   & 3     & 21  & 12    & 12       & & 24  & 12    & 84  & 48    & 48        \\ \hline
  8   & 4     & 28  & 16    & 16       & & 26  & 13    & 91  & 52    & 52        \\ \hline
  10  & 5     & 35  & 20    & 20       & & 28  & 14    & 98  & 56    & 56        \\ \hline
  12  & 6     & 42  & 24    & 24       & & 30  & 15    & 105 & 60    & 60        \\ \hline
  14  & 7     & 49  & 28    & 28       & & 32  & 16    & 112 & 64    & 64        \\ \hline
  16  & 8     & 56  & 32    & 32       & & 34  & 17    & 119 & 68    & 68        \\ \hline
  18  & 9     & 63  & 36    & 36       & & 36  & 18    & 126 & 72    & 72        \\ \hline
\end{tabular}
}
\end{center}
\end{table}

\begin{table}[h]
\begin{center}
\caption{Constant weight code $C_{(s_1,s_2,\cdots,s_5)}^{(h_1,h_2,\cdots,h_5)}$ for $k=3$.}
\addtocounter{table}{0}
\scalebox{1.2}{
\begin{tabular}{|cccccccccc|ccc|} \hline
  $s_1$ & $h_1$ & $s_2$ & $h_2$ & $s_3$ & $h_3$ & $s_4$ & $h_4$ & $s_5$ & $h_5$ & $N_1$ & $d$ & $d_{BV}$ \\ \hline
   2    & 1     & -     & -     & -     & -     & -     & -     & -     & -     & 3     & 2   & 2        \\
   2    & 1     & 3     & 1     & -     & -     & -     & -     & -     & -     & 10    & 5   & 5        \\
   2    & 1     & 3     & 2     & -     & -     & -     & -     & -     & -     & 11    & 6   & 6        \\
   2    & 1     & 3     & 1     & 2     & 2     & -     & -     & -     & -     & 22    & 11  & 11       \\
   2    & 1     & 3     & 1     & 2     & 1     & -     & -     & -     & -     & 23    & 12  & 12       \\ \hline
   2    & 1     & 3     & 1     & 2     & 2     & 2     & 2     & -     & -     & 46    & 23  & 23       \\
   2    & 1     & 3     & 1     & 2     & 1     & 2     & 1     & -     & -     & 47    & 24  & 24       \\
   2    & 1     & 3     & 1     & 2     & 1     & 2     & 1     & 2     & 2     & 94    & 47  & 47       \\
   2    & 1     & 3     & 1     & 2     & 1     & 2     & 1     & 2     & 1     & 95    & 48  & 48       \\ \hline
\end{tabular}
}
\end{center}
\end{table}

In Table 5 we show examples of constant weight code for $k \geqq 2$ and $N_1 \leqq 128$.

\begin{table}[h]
\begin{center}
\caption{Constant weight code $C_{(s_1,s_2,\cdots,s_5)}^{(h_1,h_2,\cdots,h_5)}$ for any $k$.}
\addtocounter{table}{0}
\scalebox{1.2}{
\begin{tabular}{|cccccccccc|cccc|} \hline
  $s_1$ & $h_1$ & $s_2$ & $h_2$ & $s_3$ & $h_3$ & $s_4$ & $h_4$ & $s_5$ & $h_5$ & $N_1$ & $k$ & $d$ & $d_{BV}$ \\ \hline
   2    & 1     & -     & -     & -     & -     & -     & -     & -     & -     & 3     & 2   & 2   & 2        \\
   2    & 1     & 4     & 2     & -     & -     & -     & -     & -     & -     & 14    & 3   & 8   & 8        \\
   2    & 1     & 4     & 2     & 2     & 2     & -     & -     & -     & -     & 30    & 4   & 16  & 16       \\
   2    & 1     & 4     & 2     & 2     & 2     & 2     & 2     & -     & -     & 62    & 5   & 32  & 32       \\
   2    & 1     & 4     & 2     & 2     & 2     & 2     & 2     & 2     & 2     & 126   & 6   & 64  & 64       \\ \hline
\end{tabular}
}
\end{center}
\end{table}


\newpage
\section{Conclusion} 
We have presented a new method of constructing superimposed code $C_{(s_1,s_2,\cdots,s_I)}^{(h_1, h_2, \cdots, h_I)}$ 
and have also presented a large class of optimum constant weight code.

We have presented large classes of optimum constant weight codes for $k=2$ and $k=3$, for $n \leqq 128$.

We have presented optimum linear constant weight codes that meet BV bound given for $n \leqq 128$, $k=2,3,4,5$ and 6.

\begin{itembox}[c]{{\bf Conjectures}}
(I) Code $C_{(s_1)}^{(h_1)} (k=2)$ yields the optimum sonstant weight code at any code length $N_1$ and at any minimum distance $d$ such that
$N_1=3h_1$ and $d=2h_1$ for any $h_1 \geqq 1$. \\
(\ii) The code $C_{(s_1)}^{(h_1)} (k=3)$ yields the optimum linear code at any code length $N_1$ and at any minimum distance $d$ such that 
$N_1=7h_1$ and $d = \frac{N_1+h_1}{2}$ for any $h_1 \geqq 1$, for $k=3$. \\
(\iii)There exist optimum constant weight of length $2^{k+1}-2$ and minimum distance $2^{k}$, for any $k \geqq 3$. 
\end{itembox}

\vspace*{1em}

\section*{Acknowledgement} 

The authors would like to thank Prof. M. Kobayashi at Shonan Institute of Technology 
and Prof. H. Yagi at The University of Electro-Communications, for their helpful discussions.




\begin{thebibliography}{9}

\bibitem{bib:3}
F.J.MacWilliams and N.J.A.Sloane, "The Theory of Error-Correcting Codes", 
North-Holland, (1977).

\bibitem{bib:2}
M.Kasahara and S.Hirasawa, "Construction of Superimposed Codes and the Decoding Algorithm", 
Proc. of SITA '92, pp.441-444, (1992-09).

\bibitem{bib:4}
M.Kasahara, Y.Sugiyama, S.Hirasawa, and T.Nanekawa, "New Class of Binary Codes Constructed on the Basis of Concatenated Codes and Product Codes", 
IEEE Trans. Inform. Theory, Vol.IT-22, No.4, pp.462-468, (1976-07).

\bibitem{bib:1}
A.E.Brouwer and Tom Verhoeff, "An updated Table of Minimum-Distance Bounds for Binary Linear Codes",
IEEE Trans. Inform. Theory, Vol.39, No.2, pp.662-677, (1993-03).

\bibitem{bib:5}
H.J. Helgert and R.D. Stinaff, "Minimum-distance bounds for binary codes," 
IEEE Trans. Inform. Theory, Vol. IT-19, pp.344-356, (1973-5).

\bibitem{bib:6}
M.Kasahara and S.Hirasawa, "Construction of A Large Class of Optimum Linear Codes for Small Number of Information Symbols",
IEICE Tech. Report, IT 2009-137, (2010-3).

\bibitem{bib:7}
W.W.Peterson,"Error-Correcting Codes",MIT Press,(1961).


\end{thebibliography}
\end{document}